\documentclass[twocolumn]{aastex6}
\usepackage{amsmath}

\shorttitle{CR pressure in galactic outflows}
\shortauthors{C. M. Simpson et al.}

\begin{document}
 
\newcommand{\Msun}{M$_\odot$}
\newcommand{\Lsun}{L$_\odot$}
\newcommand{\Zsun}{Z$_\odot$}
\newcommand{\degree}{\ensuremath{^\circ}}
\newcommand{\cm}{cm$^{-3}$}
\newcommand{\kms}{km~s$^{-1}$}

\title{The role of cosmic ray pressure in accelerating galactic outflows}
\author{Christine~M.~Simpson\altaffilmark{1,2}, 
	R\"udiger~Pakmor\altaffilmark{1},
	Federico~Marinacci\altaffilmark{3},
	Christoph~Pfrommer\altaffilmark{1},
	Volker~Springel\altaffilmark{1,4},
	Simon~C.~O.~Glover\altaffilmark{5},
	Paul~C.~Clark\altaffilmark{6} and
	Rowan~J.~Smith\altaffilmark{7}}

\altaffiltext{1}{Heidelberger Institut f\"{u}r Theoretische Studien, 
Schloss-Wolfsbrunnenweg 35, 69118 Heidelberg, Germany}
\altaffiltext{2}{Christine.Simpson@h-its.org}
\altaffiltext{3}{Kavli Institute for Astrophysics and Space Research, 
Massachusetts Institute of Technology, Cambridge, MA 02139, USA}
\altaffiltext{4}{Zentrum f\"ur Astronomie der Universit\"at Heidelberg, 
ARI, M\"{o}nchhofstr. 12-14, 69120 Heidelberg, Germany}
\altaffiltext{5}{Zentrum f\"ur Astronomie der Universit\"at Heidelberg, 
ITA, Albert-Ueberle-Str. 2, 69120 Heidelberg, Germany}
\altaffiltext{6}{School of Physics and Astronomy, Queen's Buildings, 
The Parade, Cardiff University, Cardiff CF24 3AA, UK}
\altaffiltext{7}{Jodrell Bank Centre for Astrophysics, 
University of Manchester, Oxford Road, Manchester M13 9PL, UK}

\begin{abstract}
  We study the formation of galactic outflows from supernova 
  explosions (SNe) with the moving-mesh code {\small AREPO} in a
  stratified column of gas with a surface density similar to the Milky
  Way disk at the solar circle.  We compare different simulation
  models for SNe placement and energy feedback, including cosmic rays
  (CR), and find that models that place SNe in dense gas and account
  for CR diffusion are able to drive outflows with similar mass loading 
  as obtained from a random placement of SNe with no CRs.  Despite 
  this similarity, CR-driven outflows differ in several other key properties 
  including their overall clumpiness and velocity.  Moreover, the forces 
  driving these outflows originate in different sources of pressure, with 
  the CR diffusion model relying on non-thermal pressure gradients to 
  create an outflow driven by internal pressure and the random-placement 
  model depending on kinetic pressure gradients to propel a ballistic 
  outflow. CRs therefore appear to be non-negligible physics in the 
  formation of outflows from the interstellar medium.\vspace*{0.1cm}
\end{abstract}


\section{Introduction}

Stellar feedback plays a critical role in galaxy- and star-formation
through its regulation of the interstellar medium (ISM) \citep{Joung2009, 
Walch2015, Martizzi2016,Girichidis2016b} and the powering of galactic 
winds \citep{Hopkins2014, Marinacci2014, Vogelsberger2014,Schaye2015}. 
The sources of stellar feedback are varied and impart different types of 
energy on different timescales, and in different environments 
\citep[e.g.][]{Agertz2013}.  SNe are a particularly important feedback source, 
and their energy likely combines with other stellar feedback effects 
(e.g.~UV radiation from young stars) in a non-linear way to impact the 
ISM \citep{Geen2015}.

The acceleration of CRs at shock fronts in supernova remnants is a
potentially crucial aspect of SNe feedback.  Observations of local SNe
remnants suggest that of order 10\% of the explosion energy is converted to
CRs \citep{Helder2012,Morlino2012,Ackermann2013}.  CR energy, once
created, does not dissipate quickly, in contrast to cooling processes that 
operate for thermal energy.  In addition, CRs are transported through both 
advection and diffusion processes.  The diffusion process in particular 
has the ability to transport significant amounts of CR energy independent 
of bulk gas motions to distances far from CR acceleration sites, thereby
creating potentially significant pressure imbalances that can drive 
large-scale gas flows.

Previous work has already demonstrated the impact of CRs in isolated and 
cosmological simulations of galaxies \citep{Jubelgas2008, Uhlig2012, 
Booth2013, Salem2014a,Salem2014b,Pakmor2016c} and in simulations of 
the ISM \citep{Peters2015,Girichidis2016a}.  The goal of this letter is to investigate
how CRs from SNe accelerate galactic outflows, and whether diffusion 
of these CRs represents the critical physical effect that explains galactic 
outflows in a regime where the star formation rate (SFR) is local and 
varying and SNe take place in dense gas.  To this end, we test a variety 
of SNe feedback and CR transport models, combined with low-temperature 
cooling and a self-consistent multiphase ISM treatment that goes beyond 
the subgrid model used in previous galactic studies of CRs with 
{\small AREPO} \citep{Pakmor2016c}.

\section{Simulation Setup}

We simulate a tall column of stratified gas intended to represent a small portion 
of a galactic disk.  The domain dimensions are 
1~kpc ~$\times$~1~kpc~$\times$~10~kpc.  We impose periodic boundaries 
along the two short axes and outflow boundaries along the long axis.  Gas 
starts the simulation in hydrostatic equilibrium with a temperature of $10^4$~K.

Gravitational forces are computed both from gas self-gravity using a tree-based 
algorithm with mixed periodic/non-periodic boundary conditions and a constant 
softening length of $\varepsilon = 0.165$~pc; and from an analytic potential 
representing the pre-existing stellar density at startup.  We assume this fixed 
stellar density field is proportional to the initial gas density $\rho_0$ for an 
assumed gas fraction of $f_g$ via Poisson's equation in a manner analogous 
to the method of \citet{Creasey2013}: 
$\nabla^2\phi = 4\pi G \rho_0 \times (f_g^{-1} - 1)$.  

The initial gas density varies with vertical height $h$ above the box 
mid-plane along the long axis, also following the setup of \citet{Creasey2013}:  
\begin{equation}
\label{eq:rho}
\rho_0(h) = \frac{\Sigma_0}{2b_0} \mathrm{sech}^2 \left( \frac{h}{b_0} 
\right),
\end{equation}
where $\Sigma_0$ is the initial gas surface density and $b_0$ is the initial 
isothermal scale height.  We choose $\Sigma_0 = 10$~\Msun~pc$^{-2}$ and 
$f_g = 0.1$, which results in $b_0 = 100$~pc.  The initial gas density of cells 
above 4.4~kpc is limited to a minimum value of $10^{-20}$~\Msun~pc$^{-3}$.  
Galactic shearing effects are neglected.

Hydrodynamics is computed to second order with the moving-mesh code
{\small AREPO} \citep{Springel2010, Pakmor2016a}.  {\small AREPO}
yields a quasi-Lagrangian solution to the ideal hydrodynamic equations
that captures shocks and discontinuities well.  We assume a thermal
adiabatic index of $\gamma = 5/3$ and impose an effective pressure
floor in the Riemann solver equal to $4^2$ times the Jeans pressure at
the minimum allowed cell diameter to provide pressure support in
under-resolved dense gas \citep{Machacek2001}. A minimum allowed
temperature of 5~K is adopted.

Initially, the simulated volume is divided into $10^6$ gas cells, concentrated 
in the mid-plane, but also comprising a Cartesian background mesh with a 
cell length of 43.5~pc up to 1~kpc and of 90.9~pc beyond.  Refinement and 
derefinement of the mesh is applied to maintain roughly constant cell masses 
to within a factor of two of the target gas mass of 10~\Msun, subject to the 
constraints that cell volumes are limited to between 2.93~pc$^3$ and $7.19 
\times 10^5$~pc$^3$; a maximum volume ratio of 10 between adjacent cells 
is maintained; and cell diameters are required to be no larger than $1/4$ 
of the Jeans length.

We use the chemistry and cooling network implemented by \citet{Smith2014}. 
This model solves hydrogen chemistry, including H$_2$ 
\citep{Glover&MacLow2007a,Glover&MacLow2007b}, and has a simple 
treatment for CO chemistry \citep{Nelson&Langer1997,Glover&Clark2012}.  
We assume the same species abundances for carbon, oxygen, helium, 
and dust as used by \citet{Smith2014} and the same initial ionization fractions 
and uniform interstellar radiation field as used in their fiducial model. Gas 
self-shielding and dust shielding are accounted for using the TreeCol 
algorithm \citep{Clark2012}.  Metal cooling of high-temperature gas assuming 
collisional ionization equilibrium is also included 
\citep{Gnat&Ferland2012,Walch2015} assuming a constant solar gas metallicity.

In most simulations, we include ideal magnetohydrodynamics (MHD)
computed with a Powell cleaning scheme \citep{Pakmor2011} for
divergence control.  We use an initial seed field with a strength of
$10^{-10}\, {\rm  G} \times {\rm sech}^{4/3}(h/b_0) $, oriented parallel to
the disk plane.  In a subset of our simulations, CRs are followed with
a two fluid approximation, assuming an adiabatic index of 
$\gamma_{\rm CR} = 4/3$ and including a CR cooling model that dissipates 
CR energy through Coulomb and hadronic processes \citep{Pfrommer2016}. 

\section{Tested Models}

\begin{figure*}
\centering 
\includegraphics[scale=0.7] {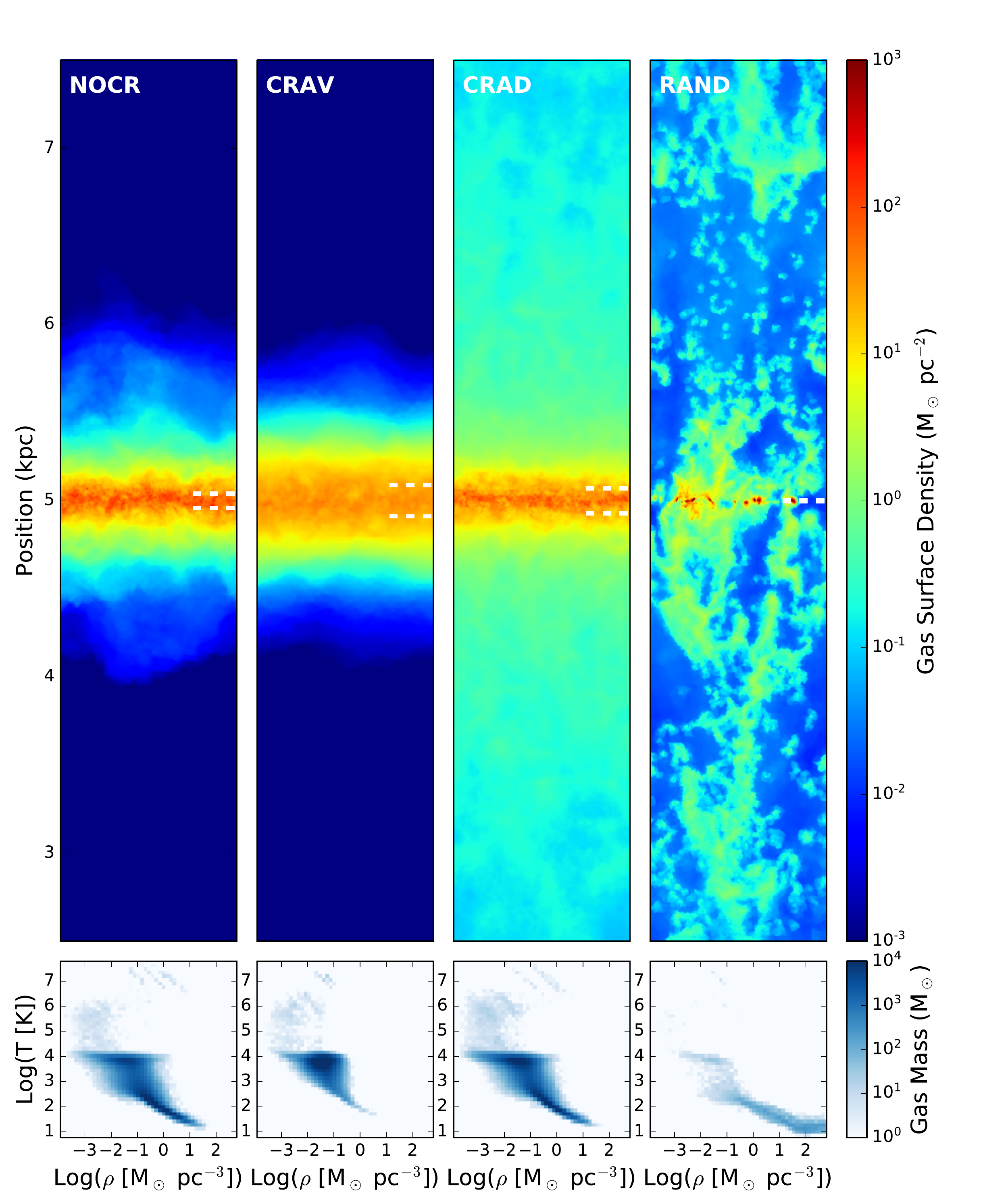}
\caption{\textit{Top row:} Unweighted projections of gas density for
  four of our models (NOCR, CRAV, CRAD, and RAND) after 100~Myr
  of evolution.  The projections show the central 5~kpc of the tall
  box and are 1~kpc wide and 1~kpc deep.  Dashed horizontal lines at
  $\pm h_{1/2}$ show the height containing half the original mass.
  \textit{Bottom row:} Gas phase space diagrams of material within 
  $\pm h_{1/2}$ of the box mid-plane.}
\label{fig:images}
\end{figure*}

We investigate several models for supernova feedback and galactic wind
acceleration.  In all models, SNe are modeled as discrete explosions of 
$10^{51}$~erg deposited into the 32 closest cells to the explosion position.  
Explosion events are only added to the mesh when all gas cells are 
synchronized; the maximum allowed timestep is 0.1~Myr.  SNe are injected 
stochastically, assuming a rate of 1.8~SNe per 100~\Msun\ of newly formed 
stars.  SNe energy is split between three energetic channels: thermal, kinetic, 
and CR.  The six models explored are as follows:

\begin{itemize}
\item In NOCR, all SNe energy is thermal and distributed over the explosion 
  cells proportional to each cell's volume.  Sites for SNe are chosen
  probabilistically, with a local SFR computed for each cell from the local 
  free-fall time, which depends on the cell's total baryon density $\rho_{b,i}$:
  $t_{{\rm ff},i} = \sqrt{3\pi/32 G \rho_{b,i}}$.  It also depends on the cell mass 
  $m_i$ and a star formation efficiency $\epsilon$, which we assume to be 0.01,
  yielding the cell's SFR:
\begin{equation}
{\rm sfr}_i = \epsilon \frac{m_i}{t_{{\rm ff},i}}.
\end{equation}
The probability of a SNe at the cell's position in a timestep $\Delta t$ is then 
computed as
\begin{equation}
p_i = {\rm sfr}_i \times \frac{1.8 \mathrm{\ SNe}}{100 \mathrm{\ M}_{\odot}} 
\times \frac{\Delta t}{m_i}.
\end{equation}
\item KE30 is identical to NOCR in the selection of SNe sites, with the only 
  difference being that 30\% of the SNe energy is added in kinetic form.  
  Explosion cells are given momenta directed radially away from the central 
  cell analogous to the method of \citet{Simpson2015}.
\item CRAV is again identical to NOCR and differs only in that 10\% of the 
  SNe energy is put into CR energy.  The remaining 90\% is added as 
  thermal energy. The CR energy can advect with the gas but no other CR 
  transport mechanism is included.
\item CRID is the same as CRAV, except that this model includes the additional 
  CR transport mechanism of isotropic diffusion, as described by \citet{Pakmor2016b}.  
  A diffusion coefficient of $\kappa = 10^{28}$~cm$^2$~s$^{-1}$ is used.
\item CRAD is also identical to CRAV, but it includes {\em anisotropic} instead of 
  isotropic CR diffusion \citep{Pakmor2016b}.  The diffusion coefficient in this 
  model is $\kappa = 10^{28}$~cm$^2$~s$^{-1}$ parallel to the magnetic field and 
  zero in all transverse directions.
\item RAND differs from all the other models in the way the locations of SNe 
  are chosen.  Rather than computing a local SFR for each cell, a global SFR 
  for the entire volume is calculated from the gas column density according 
  to the empirical Kennicutt star-formation relation \citep{Kennicutt1998}.  As 
  mass is lost from the volume, the SFR is adjusted to the new gas column 
  density.  The locations of SNe explosions are randomly distributed, 
  uniformly in the plane parallel to the disk and following the functional form of 
  Eqn.~(\ref{eq:rho}) in the vertical direction.  The scale height $b$ of the 
  latter distribution is varied according to the current height containing half the 
  initial mass of the box, $h_{1/2} = 0.55 b$.  RAND is intended to test a mode 
  of wind generation that does not rely on CR effects, but rather on decoupling 
  SNe locations from dense gas in a `random-driving' scenario.  To this end, all 
  SNe are purely thermal, and this model does not include MHD.
\end{itemize}

\section{Summary of Results}

\begin{figure*}
\includegraphics[width=1.0\textwidth] {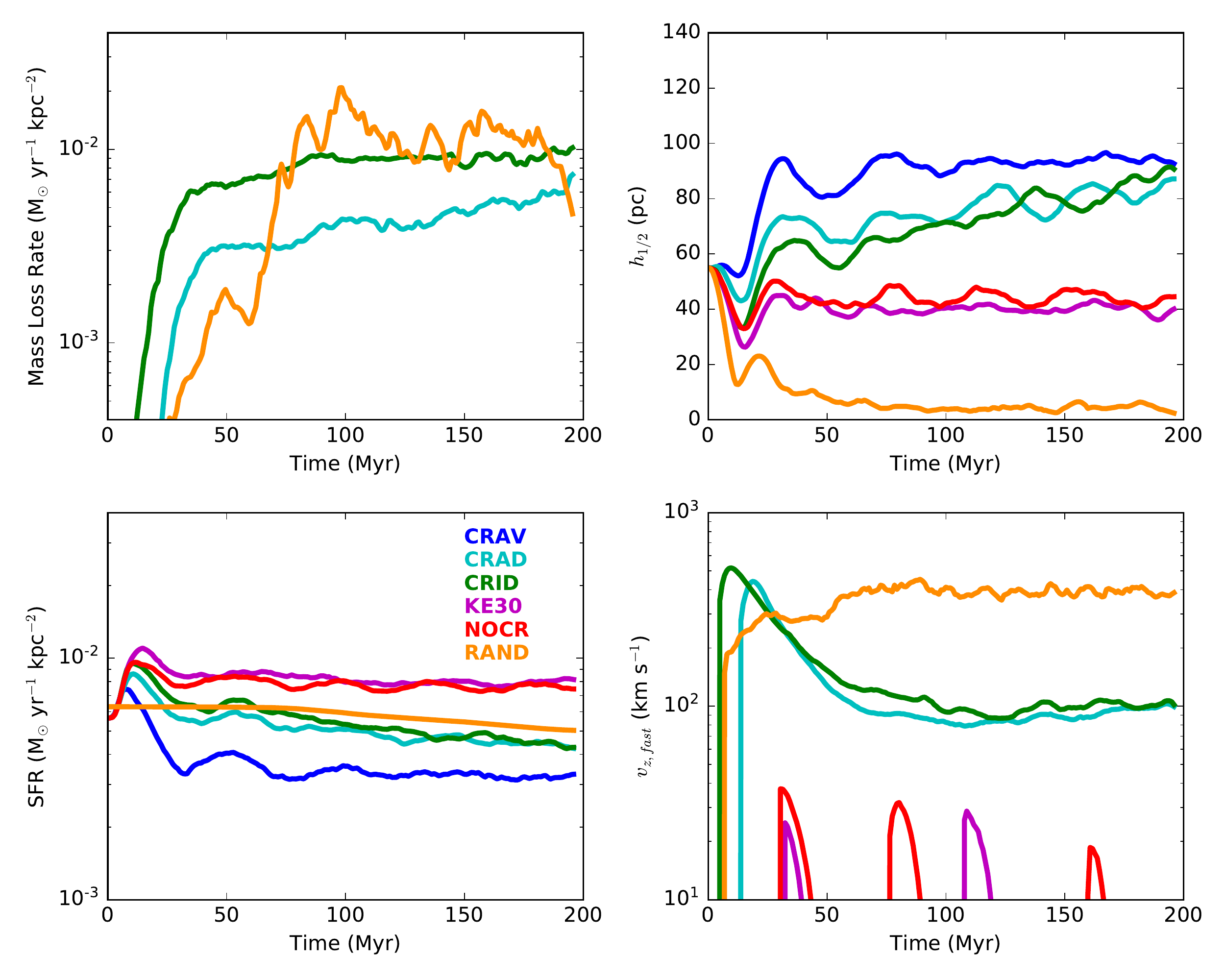} 
\caption{Time evolution of simulation properties; each model is indicated with a 
  different color.  \textit{Top left:} The mass loss rate as computed from the 
  difference in the total gas mass between successive simulation snapshots 
  separated by 1~Myr.  Only CRID, CRAD, and RAND have sufficient mass loss 
  to be included in this panel.  \textit{Top right:} Height below which the total mass 
  enclosed is half the initial mass contained within the box.  \textit{Bottom left:} 
  The total SFR.  \textit{Bottom right:} The minimum velocity of the fastest 
  $10^3$~\Msun\ of gas between 1 and 4.5~kpc from the mid-plane.}
\label{fig:evo}
\end{figure*}

A comparison of the gas density distribution after 100~Myr of evolution, shown 
in Figure~\ref{fig:images}, immediately reveals significant differences between 
several of our tested models.  The disk scale heights, mid-plane density structures, 
and extended gas distributions are all visually distinct and demonstrate key 
variations in model behavior.  In addition, the phase-space distributions of gas 
within the mid-plane (bottom panels of Figure~\ref{fig:images}) reflect the ability, 
or inability, of each model to regulate the supply of dense gas.  These differences 
also appear in the time evolution of global properties such as mass loss, SFR, and 
disk scale height, displayed in Figure~\ref{fig:evo}.

\begin{figure}
\centering 
\includegraphics[width=0.5\textwidth] {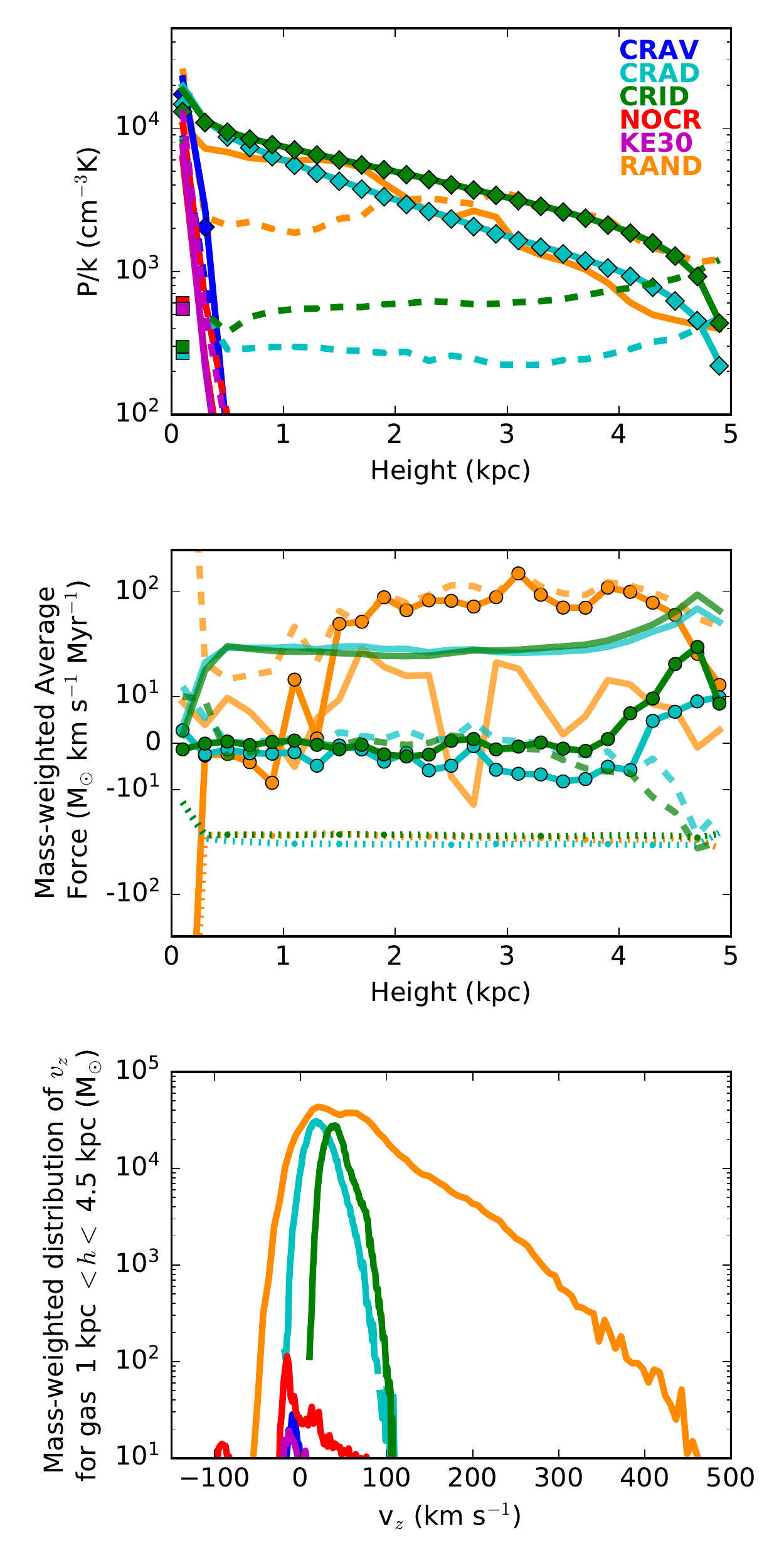}
\caption{Diagnostic quantities exploring the outflow dynamics after 100~Myr of 
  evolution.  \textit{Top panel:} Vertical profiles of the volume-weighted average 
  pressure for different pressure components: $P_{\rm int}$ (solid lines), 
  $P_{\rm kin}$ (dashed lines), $P_{\rm CR}$ (diamonds) and $P_{\rm mag}$ 
  (squares).  \textit{Middle panel:} Vertical profiles of the mass-weighted vertical 
  force for different force components: $F_{{\rm pres},z}$ (light solid lines), 
  $F_{{\rm kin},z}$ (light dashed lines), $F_{{\rm grav},z}$ (dotted lines), and 
  the sum of these three forces (solid lines with circles).  For clarity, only CRID, 
  CRAD, and CRAV are included in this panel.  \textit{Bottom panel:} 
  Mass-weighted distribution of vertical gas velocities for all gas between 1 and 
  4.5~kpc from the mid-plane.}
\label{fig:profile}
\end{figure}

In the simplest scenario tested, NOCR, where SNe are modeled as purely 
thermal energy injection events, the mid-plane gas quickly becomes a
turbulent, multi-phase medium that maintains a scale height slightly
below its initial value.  Little material in this model reaches more than a few 
hundred parsecs above the mid-plane.  Gas in the mid-plane becomes 
denser on average due to cooling, causing an overall SFR increase. 
However, the density increase is limited and regulated by SNe, which are 
preferentially injected into dense regions.

In KE30, we explore the effect of directly injecting 30\% of the SNe
energy in kinetic form.  This model produces only small differences
relative to NOCR, as shown in Figure~\ref{fig:evo}.  There appears to
be a small enhancement in the availability of dense gas in KE30, but
the overall similarity between these two models can be understood as
being primarily due to the high simulation resolution, allowing the purely 
thermal model of NOCR to closely capture the Sedov-Taylor phase of SNe 
remnants, as discussed in \citet{Simpson2015}.

Allowing the addition of CRs in SNe changes this picture significantly.  First, 
without diffusion (model CRAV), CRs have a significant impact on the 
mid-plane gas structure.  The non-thermal pressure contributed by CRs 
suppresses the formation of dense gas and increases the disk scale height.  
This results in a lower SFR. However, the new reservoir of non-thermal 
pressure is not sufficient by itself to accelerate material to significant heights
above the mid-plane.

Adding CR diffusion, as in CRAD and CRID, alters the influence of CRs
dramatically.  In these models, gas is driven from the ISM to 
significant heights above the mid-plane, yielding mass loss rates 
comparable to the SFR.  The type of diffusion also plays a 
role in the overall evolution.  The onset of diffusion-generated outflows in 
CRAD is delayed relative to CRID, and the outflows are generally weaker.  
Early in CRAD diffusion is less efficient in transporting CR energy away 
from the mid-plane because of the initial orientation of the magnetic field
parallel to the mid-plane.  This temporarily results in the trapping of CR 
energy in the mid-plane, producing a higher scale height and lower SFR, 
until the magnetic field reorients.  At late times, CRAD evolves much more 
like CRID, indicating that ISM turbulence has accomplished this and CRs 
are now able to escape the mid-plane.

Aside from the CR diffusion models, the only other scenario that produces 
robust outflows is the RAND model.  The nature of outflows between the 
CR diffusion models and RAND is quite different.  Figure~\ref{fig:images} 
shows the clumpy nature of gas above the mid-plane in RAND, contrasting 
with the much smoother flow in CRAD.
The outflows produced are also faster.  The mass loss rate of RAND is 
similar to CRID, but the SFR is larger, yielding a somewhat smaller mass 
loading of the outflow.  The mid-plane ISM in RAND undergoes a thermal 
runaway, where the mid-plane gas becomes maximally porous as most 
of the mass collapses into small, dense clumps, also seen in 
Figure~\ref{fig:images}.  The disk scale height equilibrates to approximately 
four times the minimum allowed cell diameter, implicating the imposed 
pressure floor as the main disk-support in RAND.

\section{Discussion}

\begin{figure*}
\includegraphics[width=1.0\textwidth] {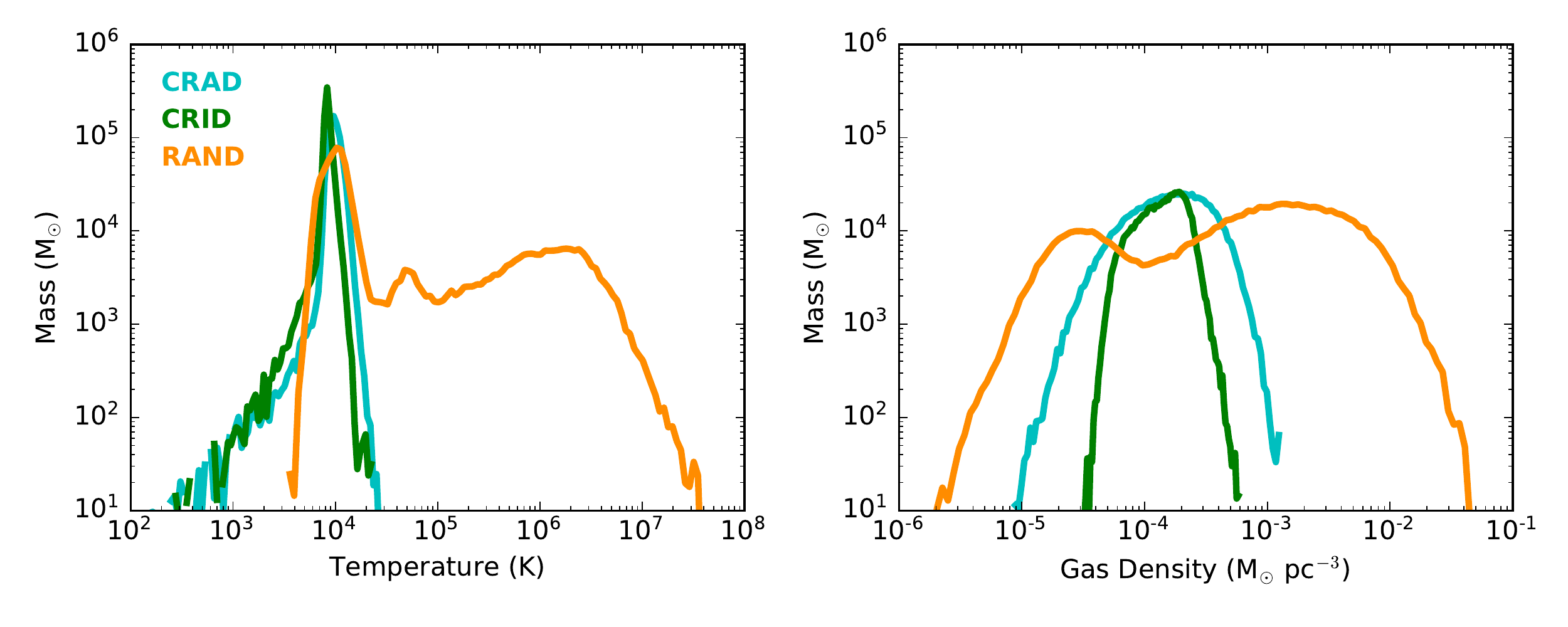} 
\caption{Mass-weighted distributions of gas temperature and density
for outflowing gas between 1~and~4.5~kpc 
above the mid-plane after 100 Myr. }
\label{fig:outflow}
\end{figure*}

Three of our tested models, CRID, CRAD, and RAND, have demonstrated
the ability to accelerate significant amounts of gas several kpc above
the mid-plane.  While the outflows in these models have similar mass loading 
factors, as revealed by Figure~\ref{fig:evo}, the physical mechanisms driving these 
flows are quite different. In fact, the simulations exhibit two distinct modes of wind 
generation: a `pressure-driven wind' and a `ballistic wind'. Figure~\ref{fig:profile} 
shows that CRID, CRAD, and RAND are the only models that have significant 
internal gas pressure at heights above 1~kpc from the mid-plane.  The internal 
pressure of cell $i$ is defined as
\begin{equation}
  P_{{\rm int},i} = (\gamma -1)\rho_i e_i + (\gamma_{\rm CR} -1)
  \rho_i e_{{\rm CR},i} + 
  \frac{B_i^2}{8\pi},
\end{equation}
where $\rho_i$ is the gas density, $e_i$ is the specific internal thermal energy, 
$e_{{\rm CR},i}$ is the specific CR energy, and $B_i$ is the magnetic field strength.  
In CRID and CRAD, the total pressure is dominated by the CR pressure term,
$P_{{\rm CR},i} = (\gamma_{\rm CR} -1) \rho_i e_{{\rm CR},i}$.  In RAND, however, 
internal pressure only dominates up to a height of 2~kpc; beyond this, the kinetic ram 
pressure, $P_{{\rm kin},i} = \rho_i v_i^2/2$, begins to dominate ($v_i$ is the gas 
speed).  The magnetic pressure, $P_{{\rm mag},i} = {B_i^2}/({8\pi})$, is subdominant 
in all models above the disk.

How do these pressures drive gas?  We consider the forces acting on the gas in the 
vertical direction $z$ to explore this question.  These forces are in effect the terms 
from the momentum-conservation equation.  They include the gravitational force:
\begin{equation}
  F_{{\rm grav},z,i} = m_i \times a_{{\rm grav},z,i},
\end{equation}
where $a_{{\rm grav},z,i}$ is the cell gravitational acceleration; the internal pressure 
force:
\begin{equation}
F_{{\rm pres},z,i} = -V_i \frac{\partial P_{{\rm int},i}}{\partial z},
\end{equation} 
where $V_i$ is the cell volume; and the kinetic force:
\begin{equation}
  F_{{\rm kin},z,i} = -\frac{V_i}{2}\frac{\partial \rho_i v_i^2}{\partial z}.
\end{equation} 

Figure~\ref{fig:profile} shows the average force acting on the gas versus 
height.  In RAND, the total force is dominated by the gravitational force within the 
disk, and by the kinetic force above the disk.  By comparison, the internal pressure 
force is not as significant and alternates between positive and negative values with 
height, likely reflecting the clumpy nature of the outflow.  In CRID and CRAD, the 
kinetic force is very small in magnitude at most heights. In contrast, the internal 
pressure and gravitational forces are more significant and of similar magnitude, but of 
opposite sign.  The gravitational force dominates on average, but as is seen in the 
distribution of gas velocities, this does not prevent individual gas elements from 
reaching high outflowing velocities, sustaining a nearly constant mass loss rate from 
the box over time.

We note that the outflows described here are unlikely to reach wind velocities large 
enough to be unbound from the galaxy.  The Milky Way escape velocity at the 
solar circle probably exceeds 500~\kms\ \citep{Smith2007}.  In addition, 
extrapolations from recent UV observations of local star-bursting galaxies suggest 
that for the SFR surface densities simulated here, outflow velocities rarely exceed 
50~\kms\ \citep{Heckman2016}.  CRID and CRAD sustain significant amounts of 
outflowing gas with velocities\footnote{The quoted outflow velocities for 
  CRID and CRAD exclude gas within 500~pc of the outflow boundaries because the 	
  CR energy of mirrored ghost cells beyond these boundaries is fixed to be zero.  This 
  gives spurious CR pressure gradients at the boundary.} above 50~\kms, but propel 
very little gas mass above 100~\kms. Similarly, RAND does produce significant 
amounts of gas above 50~\kms, but little above 500~\kms.  The `outflows' simulated 
here are therefore more accurately characterized as galactic fountain flows, which is 
also consistent with the significant amounts of gas above 1~kpc that have inflowing 
velocities in both CRAD and RAND, as shown in Figure~\ref{fig:profile}.

It is remarkable that the CR diffusion models, despite their placement of SNe in 
dense gas, produce winds of comparable mass loading to the `random-driving' 
scenario of RAND. The physical motivation for the latter is to account for plausibly 
lower SNe background densities, due to ionized H II regions around young stars or to 
`run-away' stars that move significant distances from their birth clouds before SNe 
can occur.  Global galaxy simulations suggest that in the absence of CRs this effect 
may significantly impact galaxy properties \citep{Rosdahl2015}.  How these effects 
would alter the dichotomy presented here between `pressure-driven' and `ballistic' 
winds, and alter the outflows of the CR diffusion models, will be a topic of future 
investigation.  

Our results are consistent with those of \citet{Girichidis2016a} and 
\citet{Peters2015} who also explored the role of anisotropic CR diffusion on 
galactic outflows launched from the ISM.  Both studies assumed a constant SFR 
and constant fractions of randomly-placed and clustered SNe.  Despite the different 
model for SNe placement, \citet{Girichidis2016a} found similar outflow velocities, 
suggesting that in this regime, CR diffusion may indeed be the dominant physical 
effect driving outflows.  However, \citet{Peters2015} demonstrated that the inclusion 
of self-gravity altered wind properties, suggesting some mediating role for other 
physical effects.  Both studies found that CR-driven outflows were colder and 
denser than thermally-driven outflows. Figure \ref{fig:outflow} shows that the 
outflows in RAND have two components: a hot, diffuse component, comprised of 
the high-velocity gas; and a slower, $10^4$~K-component, denser than CR-driven 
outflows of similar temperature. The adaptive nature of our mesh also gives better 
resolution in outflowing gas and may allow better resolution of density peaks in 
irregular flows.

The complex outflow in RAND is likely more sensitive to model assumptions 
such as the rate and placement of SNe than the CR-driven outflows.  A higher value for 
the global SNe rate could produce faster winds in RAND, but the value adopted for this 
rate is already greater than 100~SNe~Myr$^{-1}$, motivated by the Chabrier IMF and 
extending the mass range for core-collapse SNe-producing stars down to 6 \Msun\ 
\citep{Creasey2013}.  RAND should be considered an upper limit to the outflow-efficiency 
of purely-random thermal feedback.  We will also note that these models when applied 
to higher gas surface densities found in starbursting systems or in galactic centers may 
produce faster outflows, possibly exceeding galactic escape velocities.

Our models lack several effects potentially important for modeling CR-driven 
outflows.  CR streaming, not included here, may modify CR-driven outflows 
\citep{Ruszkowski2016} by possibly altering CR fluxes and heating thermal gas through 
the excitation of Alfv\'en waves \citep{Uhlig2012}.  Galactic shear may also be important, 
because of its impact on the magnetic field orientation and therefore on the diffusivity of 
CRs in our anisotropic scheme.  Our lack of an ordered, disk-parallel magnetic field
in energy equipartition with the thermal gas may also impact the formation of this 
instability \citep{Parker1966}, however, our small horizontal box width (1~kpc) may 
limit the fastest-growing modes of the Parker instability that typically have wavelengths 
close to this value \citep{GizShu1993,Rodrigues2016}.

In conclusion, the models presented here underline the importance of CR physics for 
driving galactic outflows.  A full understanding of the impact of these outflows on 
galaxy evolution will require self-consistent simulations on global galactic scales. The 
methods explored here make use of adaptive and individual timesteps, making these 
models more readily extendable to a variety of galactic contexts and a promising 
direction for our work.

\begin{acknowledgements}
We would like to thank the anonymous referee whose helpful comments improved this 
Letter.  This work was supported by the European Research Council under ERC-StG grant 
EXAGAL-308037 and ERC-CoG grant CRAGSMAN-646955 and by the Klaus Tschira 
Foundation.  SCOG acknowledges financial support from the Deutsche 
Forschungsgemeinschaft (DFG) via SFB 881 ``The Milky Way System" (subprojects 
B1, B2 and B8) and via SPP 1573 ``Physics of the Interstellar Medium'' (grant 
number GL 668/2-1).  PCC acknowledges the support of a consolidated grant (ST/
K00926/1) from the UK Science and Technology Funding Council, and the EU-funded 
network ``StarFormMapper" (687528) via call H2020-COMPET-2015.  RJS 
acknowledges support through the RAS Norman Lockyer Fellowship.
\end{acknowledgements}

\bibliography{letter}

\end{document}